\documentstyle[12pt]{article}

\newcommand{\beq}{\begin{equation}}
\newcommand{\eeq}{\end{equation}}

\begin{document}

\begin{titlepage}
\begin{center}
\today     \hfill    LBNL-39096\\
%~{} \hfill \\

\vskip 0.4in

{\large \bf $(S_3)^3$ Theories of Flavor}\footnote{This work was supported 
by the Director, Office of Energy Research, Office of High Energy and 
Nuclear Physics, Division of High Energy Physics of the U.S. Department of 
Energy under Contract DE-AC03-76SF00098.}

\vskip 0.4in

Christopher D. Carone\footnote{Plenary talk presented at The 5th
International Workshop on Supersymmetry and Unification of
Fundamental Interactions (SUSY-96), University of Maryland,
College Park, May 29 -- June 1, 1996.} 

\vskip 0.1in

{\em Theoretical Physics Group\\
     Lawrence Berkeley National Laboratory\\
     University of California, Berkeley, California 94720}
        
\end{center}

\vskip .4in

\begin{abstract}
I present a supersymmetric theory of flavor based on the discrete
flavor group $(S_3)^3$.  The model can account for the masses and
mixing angles of the standard model, while maintaining sufficient
sfermion degeneracy to evade the supersymmetric flavor problem. I
demonstrate that the model has a viable phenomenology and makes one
very striking prediction: the nucleon decays predominantly to $K l$
where $l$ is a {\em first} generation lepton.  I show that the
modes $n \rightarrow K^0 \bar{\nu}_e$, $p\rightarrow K^+ \bar{\nu}_e$,
and $p\rightarrow K^0 e^+$ occur at comparable rates, and could well
be discovered simultaneously at the SuperKamiokande experiment.
\end{abstract}

\end{titlepage}
%THIS PAGE (PAGE ii) CONTAINS THE LBL DISCLAIMER
%TEXT SHOULD BEGIN ON NEXT PAGE (PAGE 1)
\renewcommand{\thepage}{\roman{page}}
\setcounter{page}{2}
\mbox{ }

\vskip 1in

\begin{center}
{\bf Disclaimer}
\end{center}

\vskip .2in

\begin{scriptsize}
\begin{quotation}
This document was prepared as an account of work sponsored by the United
States Government. While this document is believed to contain correct 
information, neither the United States Government nor any agency
thereof, nor The Regents of the University of California, nor any of their
employees, makes any warranty, express or implied, or assumes any legal
liability or responsibility for the accuracy, completeness, or usefulness
of any information, apparatus, product, or process disclosed, or represents
that its use would not infringe privately owned rights.  Reference herein
to any specific commercial products process, or service by its trade name,
trademark, manufacturer, or otherwise, does not necessarily constitute or
imply its endorsement, recommendation, or favoring by the United States
Government or any agency thereof, or The Regents of the University of
California.  The views and opinions of authors expressed herein do not
necessarily state or reflect those of the United States Government or any
agency thereof, or The Regents of the University of California.
\end{quotation}
\end{scriptsize}

\vskip 2in

\begin{center}
\begin{small}
{\it Lawrence Berkeley Laboratory is an equal opportunity employer.}
\end{small}
\end{center}

\newpage
\renewcommand{\thepage}{\arabic{page}}
\setcounter{page}{1}

\section{Introduction}

The origin of flavor has been a significant puzzle in particle physics
since the discovery of the muon.   The replication of fermion generations 
and the strongly hierarchical pattern of their masses and mixing angles 
is left unexplained in the Standard Model.  In most theories of flavor, 
new symmetries are introduced at mass scales that are large compared to 
the electroweak scale.  To stabilize this hierarchy of scales, it is 
natural to work in the framework of the supersymmetric standard model
(SSM).  However, this complicates the flavor problem by introducing a 
new sector of particles whose masses and mixing angles must also be 
understood.  While no superpartner has yet been observed, the acceptable 
spectrum is constrained by low-energy processes.  Most notably, a high 
degree of degeneracy is required among the light generation squarks to 
suppress dangerous flavor-changing effects \cite{DG}, unless there is a strong
alignment of quark and squark eigenstates \cite{NS,LNS}.  The challenge 
in a supersymmetric theory of flavor is to simultaneously explain both 
the suppression of flavor changing effects from the scalar sector and the 
hierarchical pattern of the quark Yukawa couplings. 

In this talk, I will advocate imposing a discrete, gauged non-Abelian family 
symmetry on the SSM to obtain the desired degree of sfermion 
degeneracy \cite{hallmur}.  This choice is reasonable given that global 
continuous symmetries may be broken by quantum gravitational effects \cite{qg}, 
while gauged continuous symmetries may generate $D$-term contributions to 
the squark and slepton masses that are nonuniversal \cite{Dterm}.  In
Ref.~\cite{hallmur}, it was demonstrated that the non-Abelian discrete group 
$(S_3)^3$ is a promising choice for this flavor symmetry.  The group $S_3$ 
has both a doublet (${\bf 2}$) and a non-trivial singlet representation 
(${\bf 1}_A$) into which the three generations of fermions can be 
embedded. In order to construct a viable model for the quarks, three 
separate $S_3$ factors are required, for the left-handed doublet fields $Q$, 
and the right-handed singlet fields $U$ and $D$.  The first and second 
generation fields transform as doublets, which ensures the degeneracy among
the light generation squarks.  The third generation fields must then transform 
as ${\bf 1}_A$s so that the theory is free of discrete gauge anomalies.  
While the group $S_3$ acts identically on three objects, the representation
structure distinguishes between the generations.  Thus, it is possible to
choose the quantum numbers for the Higgs fields so that only the top 
Yukawa coupling is allowed in the symmetry limit.  The hierarchical 
structure of Yukawa matrices can then be understood as a consequence of the 
sequential breaking of the flavor symmetry group.  

After reviewing the basic scenario in the quark sector \cite{hallmur}, 
I show how we may extend the $(S_3)^3$ model to the leptons \cite{CHM1}.
This work was done in collaboration with Larry Hall and Hitoshi Murayama.
I assume that the fundamental sources of flavor symmetry breaking are 
gauge singlet fields $\phi$ that transform in the same way as the 
irreducible ``blocks'' of the quark Yukawa matrices.  I will call these 
fields `flavons' below.  I assume that the fermion Yukawa matrices arise 
from higher dimension  operators involving the $\phi$ fields, 
that are present at the Planck scale.  With flavor symmetry breaking 
originating only from the Yukawa matrices, I  estimate the 
contributions to lepton flavor violation and proton decay.  The latter
originates from non-renormalizable operators that conserve $R$-parity, 
but violate baryon and lepton number, that are also presumably 
generated at the Planck scale.  I show that the flavor symmetry is 
sufficient to suppress these operators to an acceptable level.  In 
addition, I show that the dominant proton decay modes in the $(S_3)^3$ 
model are of the form $p \rightarrow K l$, where $l$ is a {\em first} 
generation lepton.  This is never the case in either supersymmetric or 
non-supersymmetric grand unified theories.  The prediction of these 
rather unique modes is exciting since the total decay rate is likely to 
be within the reach of the SuperKamiokande experiment.

\section{The Basic Model}

In the $(S_3)^3$ model of Ref.~\cite{hallmur}, the quark chiral 
superfields $Q$, $U$, and $D$ are assigned to ${\bf 2}+{\bf 1_A}$ 
representations of $S_3^Q$, $S_3^U$ and $S_3^D$, respectively. The 
first two generation fields are embedded in the doublet, for the reasons
described in the Introduction.  The Higgs fields both transform 
as $({\bf 1}_A, {\bf 1}_A, {\bf 1}_S)$'s, so that the top quark Yukawa 
coupling is invariant under the flavor symmetry group.  The
transformation properties of the Yukawa matrices are:
\beq
Y_U \sim \left(
\begin{array}{cc|c} 
\multicolumn{2}{c|}{
({\bf \tilde{2}},{\bf \tilde{2}},{\bf 1}_S) }
& ({\bf \tilde{2}},{\bf 1}_S,{\bf 1}_S) \\ \hline
\multicolumn{2}{c|}{({\bf 1}_S,{\bf \tilde{2}},{\bf 1}_S)} &
({\bf 1}_S,{\bf 1}_S,{\bf 1}_S) \end{array} \right) 
\eeq
\beq
Y_D \sim \left(\begin{array}{cc|c} 
\multicolumn{2}{c|}{
({\bf \tilde{2}},{\bf 1}_A,{\bf 2})}
& ({\bf \tilde{2}},{\bf 1}_A,{\bf 1}_A) \\ \hline
\multicolumn{2}{c|}{({\bf 1}_S,{\bf 1}_A,{\bf 2})} &
({\bf 1}_S,{\bf 1}_A,{\bf 1}_A) \end{array}\right) 
\label{eq:transp}
\eeq
where I use the notation ${\bf \tilde{2}} \equiv {\bf 2} \otimes {\bf
1}_A$,\footnote{${\bf \tilde{2}} = (a,b)$ is equivalent to ${\bf 2} =
(b, -a)$.}. Note that these matrices involve at most 7 irreducible multiplets
of $(S_3)^3$.  In Ref.~\cite{hallmur}, $(S_3)^3$ was broken by only four types
of flavons: $\phi({\bf\tilde{2}}, {\bf 1}_S, {\bf 1}_S)$, $\phi({\bf
\tilde{2}}, {\bf\tilde{2}}, {\bf 1}_S)$, $\phi({\bf 1}_S, {\bf 1}_A, {\bf
1}_A)$, and $\phi({\bf \tilde{2}}, {\bf 1}_A, {\bf 2})$, the minimal number
which leads to realistic masses and mixings \cite{CM3}:
\beq
Y_U = \left( \begin{array}{cc|c}
      h_u & h_c \lambda & - h_t V_{ub}\\ 
      0 & h_c & - h_t V_{cb} \\ \hline 
      0 & 0 & h_t 
            \end{array} \right) ,
\eeq
\beq
Y_D = \left( \begin{array}{cc|c}
      h_d & h_s \lambda & 0\\ 
       0 & h_s & 0 \\ \hline 0 & 0 & h_b
            \end{array} \right),
\eeq
with $\lambda \simeq 0.22$.  The form of the Yukawa matrices presented 
above can be understood as a consequence of a sequential breaking of 
the flavor symmetry.  I will assume that the $2\times 2$ blocks 
of $Y_U$ and $Y_D$ are each generated by two flavon fields that 
acquire vevs at different stages of the symmetry breaking.  Thus,
$Y_D = Y_1 + Y_2$, and $Y_U=Y'_1+Y'_2$ where
\begin{equation}
Y_1 = \left( \begin{array}{cc}
      0 & a h_s \lambda \\ 0 & h_s    
\end{array} \right) \,\,\, , \,\,\,
Y_2 = \left( \begin{array}{cc}
      h_d & 0 \\ 0 & 0
      \end{array} \right).
\label{eq:y}
\end{equation}
and
\begin{equation}
Y'_1 = \left( \begin{array}{cc}
      0 & a' h_c \lambda \\ 0 & h_c
\end{array} \right) \,\,\, , \,\,\,
Y'_2 = \left( \begin{array}{cc}
      h_u & 0 \\ 0 & 0
      \end{array} \right).
\label{eq:y'}
\end{equation}
Note that $a$ and $a'$ are order one constants, with $a-a'=1$.

\section{Incorporating Lepton Sector}
Three principles determine the precise transformation 
properties of the lepton fields:
\begin{enumerate}
\item  There are no new flavor symmetries (e.g. new $S_3$ factors)
that arise only in the lepton sector. The only flavor symmetry in 
the theory is $S_3^Q \times S_3^U \times S_3^D$.
\item  The transformation properties of the lepton fields are chosen 
so that the charged lepton Yukawa matrix is similar to that of the
down quarks. 
\item  The most dangerous dimension-five operator
that contributes to proton decay, $(QQ)(QL)$, is forbidden in the $(S_3)^3$ 
symmetry limit.
\end{enumerate}
As we will see below, these principles are sufficient to completely 
determine the transformation properties of the lepton fields.

Let us first consider the consequences of the first two conditions.
The down-quark Yukawa matrix is a coupling between the left-handed quark
fields $Q \sim ({\bf 1}_A+{\bf 2}, {\bf 1}_S, {\bf 1}_S)$ and the right-handed
down quark fields $D \sim ({\bf 1}_S, {\bf 1}_S, {\bf 1}_A + {\bf 2})$.
We know that the Yukawa matrix of the charged leptons is quite similar
to that of the down quarks, up to factors of order
three \cite{GJ} at high scales:
\begin{equation}
m_b \simeq m_\tau, \,\,\,\,\,\,\,\,\, m_s \simeq \frac{1}{3} m_\mu,
\,\,\,\,\,\,\,\,\, m_d \simeq 3 m_e.
\label{eq:3factors}
\end{equation}  
Therefore, we look for an assignment of lepton transformation properties
that leads automatically to this observed similarity.  There are only 
two possibilities:
\begin{equation}
\begin{array}{c|cc}
& S_3^Q & S_3^D \\ \hline L & {\bf 1}_A+{\bf 2} &
{\bf 1}_S \\ E & {\bf 1}_S & {\bf 1}_A + {\bf 2}
\end{array}
\mbox{ or }
\begin{array}{cc}
S_3^Q & S_3^D\\ \hline {\bf 1}_S & {\bf
1}_A+{\bf 2} \\ {\bf 1}_A + {\bf 2} & {\bf 1}_S
\end{array}
\label{true}
\end{equation}
The third condition above allows us to distinguish between these
two alternatives. In the first assignment, the operator 
$(Q_i Q_i) (Q_j L_j)$ is allowed by the $(S_3)^3$ symmetry, and
we have proton decay at an unacceptable rate.  Therefore, only the second 
assignment in Eq.~(\ref{true}) satisfies all three criteria listed 
above. 

The remaining question that we need to answer is how the factors of 
three in Eq.~(\ref{eq:3factors}) enter in the Yukawa matrices.  
One plausible explanation is that they originate from fluctuations
in the order one coefficients that multiply the $(S_3)^3$ breaking
parameters.  Thus, we assume $Y_D=Y_1+Y_2$, while 
$Y_l = 3 Y_1 + \frac{1}{3} Y_2$.

\subsection{Lepton Flavor Violation}

The strongest constraint on lepton flavor violation comes from the
non-observation of the $\mu \rightarrow e \gamma$ decay mode.  In our
model, the contribution of the off-diagonal term in the purely 
left-handed slepton mass matrix (the LL matrix) is small 
enough ($\sim h_s^2 \lambda \sim 1 \times 10^{-7}$) to avoid the experimental
constraint for any value of $m_{\tilde{l}}$ above the LEP bound.  The
stringent limits come from the purely right-handed slepton mass matrix (RR)
and the left-right (LR) matrix, which we discuss in this section.  For
simplicity, we work in the approximation where the exchanged neutralino is a
pure bino state.

The one-loop slepton and bino exchange diagram that picks up the off-diagonal
(2,1) component in LR mass matrix generates the operator
\begin{equation}
\frac{e}{2} F_2(M_1^2,(m_{LR}^2)_{21},m^2_R, m^2_L)
      \,\,\, \bar{e}_R i\sigma^{\mu\nu} \mu_L
      F_{\mu\nu} \, , 
\label{eq:mego}
\end{equation}
where $F_{\mu\nu}$ is the electromagnetic field strength and the
function $F_2$ is defined in Ref.~\cite{CHM1}.

The decay width is given by 
\beq
\Gamma( \mu \rightarrow e \gamma) = \frac{\alpha}{4} m_\mu^3 |F_2|^2 \,, 
\eeq
and the bound Br$( \mu \rightarrow e \gamma) < 4.9 \times 10^{-11}$ implies
$|F_2| < 2.6 \times 10^{-12}$~GeV$^{-1}$.  In order to compare this bound to
the prediction of our model, let us take $m_R = m_L = m = 300$~GeV and 
$M_1 = 100$~GeV as a representative case.  We obtain
\begin{equation}
\frac{(m^2_{LR})_{21}}{m^2} < 1.0 \times 10^{-5}
\label{eq:cons}
\end{equation}
for this choice of parameters.  In our model, the (2,2) and (1,2)
elements in $Y_l$ belong to the same irreducible multiplet, and
diagonalization of $Y_l$ also diagonalizes LR mass matrix at $O(h_s
\lambda)$.  The term which may not be simultaneously diagonalizable
comes from the piece $Y_2 \sim h_d$, and hence
\begin{equation}
(m^2_{LR})_{21} \sim m_d \lambda A,
\end{equation}
where $m_d$ is the down quark mass evaluated at the Planck scale $m_d
\simeq 10~\mbox{MeV}/ 3$, and $A$ is a typical trilinear coupling. If 
we take $A \sim 100$~GeV, then $(m^2_{LR})_{21}/m^2 \sim 0.8 \times
10^{-6}$ and the constraint (\ref{eq:cons}) is easily satisfied.    
The (1,2) element in the LR mass matrix contributes 
in exactly the same way as the (2,1) entry, except that the chiralities
of the electron and muon in Eq.~(\ref{eq:mego}) are flipped.  Hence,
the (1,2) element is subject to the same constraint, which again is
clearly satisfied in our model.

The RR mass matrix also contributes to the operator in (\ref{eq:mego}). 
For the bino and slepton masses chosen earlier, we 
obtain the bound
\begin{equation}
\frac{(m^2_{RR})_{12}}{m^2} < 0.023,
\label{eq:mrrc}
\end{equation}
while in our model  
\begin{equation}
(m^2_{RR})_{12}/m^2 \simeq h_t V_{cb} \lambda \sim 0.009  \,\,.
\end{equation}
Thus, the bound on the (1,2) element of the RR matrix also satisfied.
Note that the contributions to $\mu\rightarrow e\gamma$ involving
mixing to the third generation scalars are comparable to those  
in the minimal SO(10) model \cite{BHS}, and hence they are 
phenomenologically safe.

\subsection{Proton Decay}

Since we have assumed that all possible
nonrenormalizable operators are generated at the Planck scale,
the task of studying proton decay in our model is a simple one.
We first write down all the possible dimension-five operators 
that contribute to proton decay and identify their transformation 
properties under $(S_3)^3$.  The coefficients can be estimated as the
product of Yukawa couplings that will produce the desired symmetry
breaking effect.  A list of possible operators and their
coefficients is given in Ref.~\cite{CHM1}.  

The leading contribution to proton decay comes from the operator  
$$
(Q_i Q_i)(Q_i L_i)/ M_*
,$$ 
where $M_* \equiv M_{Pl}/\sqrt{8\pi}$ 
is the reduced Planck mass, and where parentheses indicate a contraction
of SU(2) indices.  This operator transforms as 
a $({\bf 2}, {\bf 1}_S, {\bf 2})$ under $(S_3)^3$, and therefore has a 
coefficient of order $h_b h_s$.   The coefficients for all four components of
this  operator are
given by
\[
\frac{c}{2} \frac{h_b}{M_*} \left[ h_s (Q_2 Q_2) (Q_1 L_1)
            - h_s \lambda (Q_1 Q_1) (Q_2 L_1) \right.  \]\begin{equation}
            -  \left. {\cal O}(h_d) (Q_2 Q_2)(Q_1 L_2)
            + h_d (Q_1 Q_1) (Q_2 L_2) \right]  
\label{dimen5}
\end{equation}
where $c$ is an unknown coefficient of ${\cal O}(1)$.  The striking feature 
of this multiplet is that the operators involving first generation lepton 
fields $L_1$ have larger coefficients than those involving second generation 
fields $L_2$.  Thus, our model favors proton decay to $\nu_e$ and $e$ 
over decay to $\nu_\mu$ and $\mu$.  This result is in striking contrast 
to the situation in grand unified theories, where the amplitude is
proportional to the Yukawa coupling of the final-state lepton.
In our case, however, there is a residual $Z_2$ flavor symmetry in the 
limit where the first generation Yukawa couplings are set to zero.
Under this symmetry, the $U$, $D$, and $L$ fields of the second and 
third generation are odd, while all other matter fields are even.  
Thus, when the first generation Yukawa couplings are set to zero,  
dimension-five operators of the form $QQQL$ containing $L_2$ or $L_3$ 
are forbidden by this $Z_2$, while those involving $L_1$ are invariant. 

The predicted nucleon decay modes are obtained from Eq.~(\ref{dimen5})
by `dressing' the two-scalar-two-fermion operators with wino 
exchange. We obtain\footnote{In the
following discussion, we assume that the Cabibbo mixing originates from
the down sector, i.e. $a=1$, $a'=0$ in eq.~(\ref{eq:y}) and (\ref{eq:y'}).  
However we checked that all the results remain the
same even when the Cabibbo mixing comes from both the down and the up
sectors, or even solely from the up sector.}
\begin{eqnarray}
{\cal L} & = & \frac{\alpha_W}{2\pi} \frac{c h_b h_s}{M_*} \left[
       - \lambda (su) (d\nu_e) + \lambda (su) (ue) \right] \nonumber \\
       & \times & ( f(c,e) + f(c, d) ) \,\, ,
\label{eq:sumres}
\end{eqnarray}
where parentheses now indicate the contraction of spinor indices.
All the fields in (\ref{eq:sumres}) are in the mass eigenstate basis, and
terms of higher orders in $\lambda$ have been neglected.  The function
$f$ is the ``triangle diagram factor'' \cite{NA}, a function of the
wino and scalar masses; above we have used the fact $f(c,d) = f(u,d)$ 
to good accuracy.  The ratios of decay widths can be estimated using 
the chiral Lagrangian technique \cite{CWH,CD,HMY}.  We find
\[
\Gamma(p\rightarrow K^+ \bar{\nu}_e) : \Gamma(p\rightarrow K^0 e^+)
      : \Gamma(n \rightarrow K^0 \bar{\nu}_e) \]\begin{equation}
= 0.4 : 1 : 2.7 \,\,\, . 
\label{eq:ratios}
\end{equation}
Note that the proton's charged lepton decay mode dominates over the 
neutrino mode.  This is a consequence of the cancellation of the
operators $(du)(s\nu_e)$ between the first two terms of Eq.~(\ref{dimen5}).  
The dominance of $p\rightarrow K^0 e^+$ over 
$p\rightarrow K^+\overline{\nu}_e$ is rarely the case in SUSY-GUTs.

Finally, we come to the overall rate. We find  
{\samepage
\[
\tau(n \rightarrow K^0 \bar{\nu}_e) 
=  \frac{4 \times 10^{31} \mbox{ yrs}}{c^2} \left(  
 \frac{0.003\mbox{ GeV}^3}{\xi} \right. \]\begin{equation} 
\left. \;\;\;
\frac{0.81}{A_S} 
\frac{5}{(1+\tan^2 \beta)} \,
\frac{\mbox{TeV}^{-1}}{f(c,e)+f(c,d)} \right)^2 \, .
\label{eq:rate}
\end{equation}}
This result includes the effect of running the dimension-five operator 
between the Planck scale and $m_Z$, (a factor of $A_S = 0.81$ in the 
amplitude if $m_t = 175$~GeV, $\tan \beta = 2$, $\alpha_s (m_Z) = 0.12$) 
and between $m_Z$ and $m_n$ (a factor of $0.22$ in the amplitude).  In 
the expression above, $\xi$ is the hadronic matrix element of the 
four-fermion operator evaluated between nucleon and kaon states; its exact 
value is rather uncertain, but is estimated to be within the range 
$\xi = 0.003$--0.03~GeV$^3$.  If we take that $M_2 \sim
100$~GeV and $m_{\tilde{q}} \sim 700$~GeV, and $m_{\tilde{l}} \sim 300$~GeV,
then the triangle functions $f(c,e)+f(c,d) \sim (1.8\mbox{ TeV})^{-1}$.
Thus, if $c=1$ and $\xi=0.003$ GeV$^3$, we obtain a mean lifetime
$12.7\times 10^{31}$ years, which can be compared to the experimental bound, 
$\tau(n \rightarrow K^0 \bar{\nu}_e) > 8.6 \times 10^{31}$~years.  
It is interesting to note that the coefficient $4 \times 10^{31}$
in (\ref{eq:rate}) would be the same in the minimal SU(5) SUSY-GUT
with an extremely large color-triplet Higgs mass $M_{H_C} = 10^{17}$~GeV.  
Thus, the rate in our model is roughly comparable.  Overall, 
the $(S_3)^3$ symmetry gives us just enough suppression of dimension-five 
operators to evade the current bounds, so the model is phenomenologically 
viable.  Since the SuperKamiokande experiment is expected to extend 
Kamiokande's current reach by another factor or 30, there is a very good 
chance that the $n\rightarrow K^0\overline{\nu}_e$  mode may be seen.  It 
is an exciting prediction of this model that the $p \rightarrow K^0 e^+$ 
and $K^+ \bar{\nu}_e$ modes are likely to be seen at the same time 
because their rates are close to each other, as we saw 
in eq.~(\ref{eq:ratios}).

\section{Conclusions}
I have shown that the discrete flavor group $(S_3)^3$ can
account for the masses and mixing angles of the standard model
while simultaneously solving the supersymmetric flavor changing
problem.  The most striking prediction that emerged 
from the analysis is the dominance of proton decay to final states 
involving first generation lepton fields, unlike the case in SUSY GUTs.  
I showed that the ratios of decay widths for the largest modes 
$n \rightarrow K^0 \bar{\nu}_e$, $p\rightarrow K^+ \bar{\nu}_e$, 
and $p\rightarrow K^0 e^+$  are approximately  0.4 :: 1 :: 2.7.   Given 
the estimate of the total rate, I pointed out that all three modes may 
be within the reach of the SuperKamiokande experiment and could well be 
discovered simultaneously.

\begin{center}               
{\bf Acknowledgments} 
\end{center}
I am grateful to Lawrence Hall and Hitoshi Murayama for an enjoyable
collaboration. {\em This work was supported by the Director, Office of 
Energy Research, Office of High Energy and Nuclear Physics, Division of 
High Energy Physics of the U.S. Department of Energy under 
Contract DE-AC03-76SF00098.}
                                  
% '99' is at least as wide as the widest bibliography label, 
% could use '9' if there are less than 10 references. 

\end{document}